\documentclass[runningheads]{llncs}
\usepackage{graphicx,amssymb,amsmath,multirow,arydshln,color}

\begin{document}
\title{A Unified Hyper-GAN Model for Unpaired Multi-contrast MR Image Translation}
\titlerunning{Unpaired Multi-contrast MR Image Translation}
\author{Heran Yang\inst{1,2} \and
	    Jian Sun\inst{1,2} \and
	    Liwei Yang\inst{1} \and 
	    Zongben Xu\inst{1,2}}
%
%
\authorrunning{H. Yang et al.}
\institute{School of Mathematics and Statistics, Xi'an Jiaotong University, China \email{jiansun@xjtu.edu.cn}
	\and Pazhou Lab, Guangzhou, China
}
\maketitle  
\begin{abstract}
	Cross-contrast image translation is an important task for completing missing contrasts in clinical diagnosis.
	However, most existing methods learn separate translator for each pair of contrasts, which is inefficient due to many possible contrast pairs in real scenarios.
	In this work, we propose a unified Hyper-GAN model for effectively and efficiently translating between different contrast pairs. 
	Hyper-GAN consists of a pair of hyper-encoder and hyper-decoder to first map from the source contrast to a common feature space, and then further map to the target contrast image.  
	To facilitate the translation between different contrast pairs, contrast-modulators are designed to tune the hyper-encoder and hyper-decoder adaptive to different contrasts.  
	We also design a common space loss to enforce that multi-contrast images of a subject share a common feature space, implicitly modeling the shared underlying anatomical structures.
	Experiments on two datasets of IXI and BraTS 2019 show that our Hyper-GAN achieves state-of-the-art results in both accuracy and efficiency, e.g., improving more than 1.47 and 1.09 dB in PSNR on two datasets with less than half the amount of parameters.
\keywords{Multi-contrast MR \and Unpaired image translation  \and Unified hyper-GAN.}
\end{abstract}
\section{Introduction}
\label{sec:introduction}
Magnetic resonance (MR) imaging has been widely utilized in clinical diagnosis,
as it has a range of imaging contrasts and largely increases the diversity of diagnostic information.
However, due to practical limits, e.g., long scan time~\cite{vranic2019}, image corruption~\cite{zaitsev2015}, etc., it is often hard to collect all multi-contrast MR images of one subject. To solve this problem, a large variety of synthesis methods~\cite{bui2020flow,dar2019,huang2017,huang2020mcmt,jog2017,liu2021dual,roy2016,yu2019,yu2020sample} try to synthesize missing contrast from the available contrast, and most of them are one-to-one cross-contrast synthesis methods, i.e., one model is trained for each specific pair of contrasts.
For example, Dar et al.~\cite{dar2019} proposed a conditional generative adversarial network to translate between T1w and T2w images.
However, it is impractical to train each network for each pair of contrasts due to a range of commonly used MR contrasts in real scenarios.
In this work, we focus on tackling the multi-contrast MR image translation problem by a more efficient and  effective way  in an unpaired training setting, i.e., the training  multi-contrast MR images are not required to be captured from same subjects.

There are already several unpaired multi-contrast image translation methods in literature, e.g.,~\cite{anoosheh2018,choi2018,hui2018,sohail2019unpaired}, in recent years.
Firstly, CycleGAN~\cite{zhu2017}, as a one-to-one cross-contrast synthesis method, could be extended to learn multi-contrast mappings and require $N \times (N-1)$ generators for $N$ contrasts, which is impractical in real scenarios.
Furthermore, ComboGAN~\cite{anoosheh2018} and DomainBank~\cite{hui2018} decouple generator into encoder/decoder and reduce the requirement to $N$ contrast-specific encoders/decoders for $N$ contrasts, while the parameter size and training time of them scale linearly with the contrast number.
In addition, StarGAN~\cite{choi2018} and SUGAN~\cite{sohail2019unpaired} share a generator and discriminator for all contrasts, and rely on a contrast indicator to specify the desired output contrast.  
However, since the contrast indicator is simply concatenated with input image, it might be insufficient to control the translation process~\cite{alharbi2019,isola2017}.
In summary, existing methods either depend on multiple encoders/decoders, or insufficiently control/modulate the generator to be adaptive to contrast.

In this work, we aim to design a unified deep network  for unpaired multi-contrast MR image translation in an efficient and effective way. We design shared encoder and decoder to translate between different contrast pairs based on a common feature space constraint, and the encoding and decoding processes are respectively modulated by the source and target contrast codes. This is inspired by MR imaging~\cite{bernstein2004} that multi-contrast MR images are determined by human intrinsic tissue and scanner imaging parameters. The common feature space implicitly models the intrinsic tissue parameters, and the scanner imaging parameters are encoded to modulate the encoder and decoder in our network design. 

Specifically, we first define contrast-specific information as one-hot code indicating MR contrast, and construct two contrast modulators as hyper-network~\cite{ha2017} to respectively modulate the encoder and decoder to be adaptive to different MR contrasts. 
To enforce the common feature space shared by different contrasts, we further design a common space loss to enforce extracted deep features from different contrasts  within a common feature space, implicitly modeling the shared underlying  anatomical structures, besides  traditional adversarial~\cite{goodfellow2014} and cycle-consistency~\cite{zhu2017} losses. This unified multi-contrast MR image translation model, dubbed \textit{Hyper-GAN}, can effectively and efficiently  translate between different contrast pairs using a single network. 
Experiments on two multi-contrast brain MR datasets of IXI and BraTS 2019 show that Hyper-GAN achieves start-of-the-art results in both accuracy and efficiency, e.g., improving more than 1.47 and 1.09 dB in PSNR  on two datasets with less than half the amount of parameters.

\begin{figure}[!tp]
	\centering
	\includegraphics[width=0.75\textwidth]{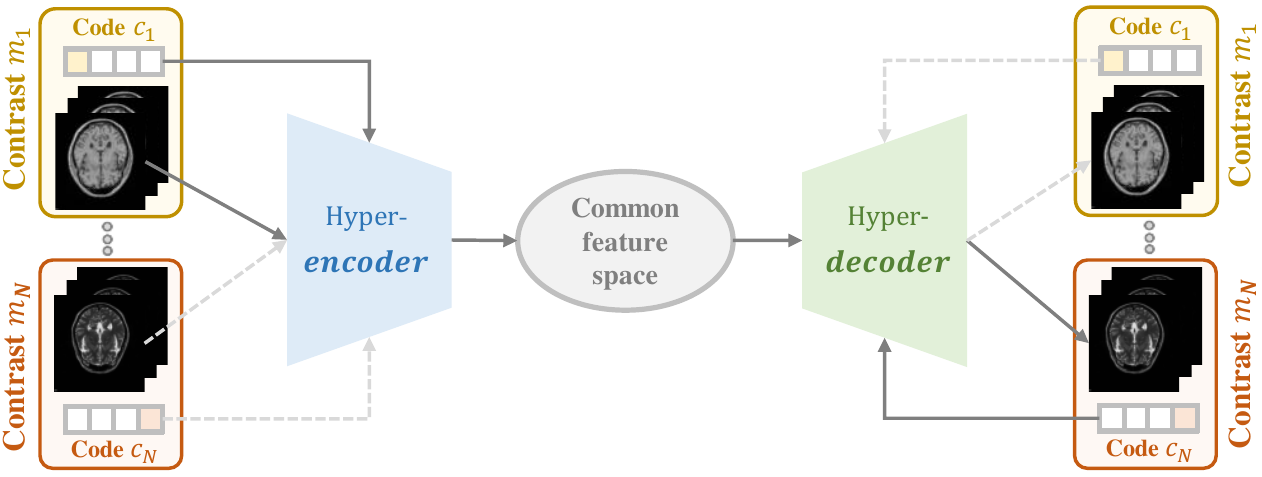}
	\caption{Hyper-GAN utilizes shared hyper-encoder and hyper-decoder
		to translate between different contrast pairs, and the encoding and decoding processes are respectively modulated by the source and target contrast codes. Multi-contrast images of a subject are constrained to be within a common feature space after encoding.} 
	\label{Fig:framework}
\end{figure}

\section{Method}
As shown in Fig.~\ref{Fig:framework}, our \textit{Hyper-GAN} utilizes a shared pair of \textit{hyper-encoder} and \textit{hyper-decoder} to translate between different contrast pairs. The source and target contrasts are each represented by a one-hot code, dubbed \textit{contrast code}, with the value of 1 representing the corresponding contrast in a given list of multiple contrasts. 
These codes are further utilized to adaptively tune the encoding and decoding processes in our hyper-encoder and hyper-decoder. Specifically, our hyper-encoder/hyper-decoder is respectively an encoder/decoder with parameters modulated by a \textit{contrast modulator} with contrast code as input, in order to make the hyper-encoder/hyper-decoder adaptive to different contrasts. 
The extracted deep features by hyper-encoder are constrained to be within a common feature space shared by different contrast images of a  subject. In this way, our Hyper-GAN is an encoder-decoder network adaptive to different source and target contrast pairs. We next introduce the details.

\subsection{Network Architecture}
\label{sec:net_arch}

\begin{figure}[!tp]
	\includegraphics[width=\textwidth]{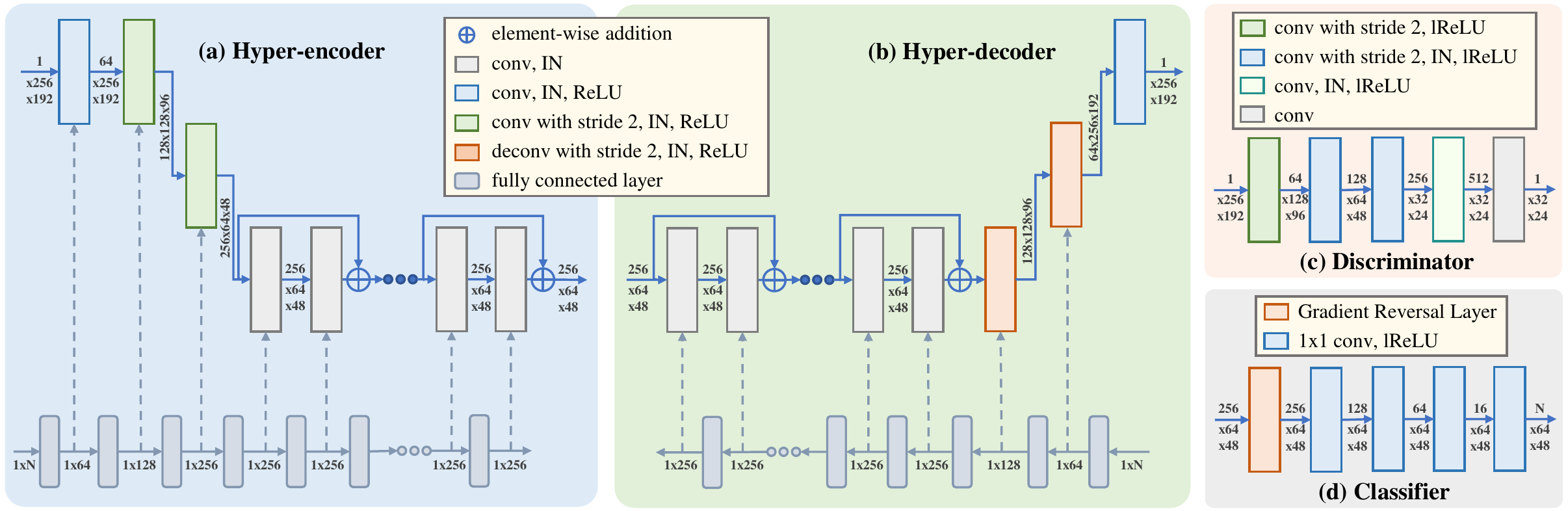}
	\caption{The illustration of the detailed network architecture in Hyper-GAN.} 
	\label{Fig:network-arch}
\end{figure}

As shown in Fig.~\ref{Fig:network-arch}(a) and (b), hyper-encoder $E$ and hyper-decoder $G$ respectively consist of two subnets, i.e., a backbone encoder/decoder and a contrast modulator.
The encoder extracts deep features from a source contrast image, while the decoder estimates the target contrast image from the extracted features. 
The encoder and decoder are respectively paired with a contrast modulator, achieving contrast-adaptive tuning of parameters of encoder and decoder based on the source and target contrast codes.
With the contrast code as input, the contrast modulator tunes the parameters using the following two different strategies, i.e., filter scaling or conditional instance normalization.

\noindent\textbf{Filter scaling (FS)}~\cite{alharbi2019}.
For each convolutional filter $f$ in encoder/decoder, the modulator produces a corresponding scalar $\alpha$ based on the contrast code and modifies this filter as $ f' \triangleq \alpha * f $, where $*$ is scalar multiplication operation.

\noindent\textbf{Conditional instance normalization (CIN)}~\cite{dumoulin2016}.
For each instance normalization (IN) layer in encoder/decoder, the modulator estimates its affine parameters $\gamma'$ and $\beta'$ based on  contrast code, and then IN layer becomes $ y = \gamma' \frac{x-\mu(x)}{\sigma(x)} + \beta'$, where $\mu(x)$ and $\sigma(x)$ are mean and standard deviation of the input features $x$.

\textbf{\textit{Network architecture}}. Encoder and decoder are respectively taken as the first five and the remaining four residual blocks of the generator in~\cite{zhu2017}, and contrast modulators are set to multilayer perceptrons. Please refer to Fig.~\ref{Fig:network-arch}(a) and (b) for the detailed definitions of hyper-encoder and hyper-decoder.
As shown in Fig.~\ref{Fig:network-arch}(c), discriminators, as in \cite{zhu2017}, are fully convolutional networks~\cite{long2015} with five $4\times4$ convolutional layers to classify whether $70\times70$ overlapping image patches are real or synthetic. 
The contrast-classifier is composed of a gradient reversal layer~\cite{ganin2015} and four $1\times1$ convolutional layers to classify which contrast the  features extracted by hyper-encoder belong to and outputs contrast probabilities.

\subsection{Training Loss}
\label{sec:train_loss}

For a $N$-contrast translation task  with contrasts $\{ {m}_k \}_{k=1}^N$ and one-hot contrast codes $\{ c_{k } \in \{0, 1\}^N\} _{k=1}^N$, our Hyper-GAN consists of a pair of hyper-encoder $E$ and hyper-decoder $G$, a contrast-classifier $C$ and contrast-specific discriminators $\{ D_{k} \}_{k=1}^N$. 
Note that $C$ and $\{ D_k \}_{k=1}^N$ are only utilized in training.

Our Hyper-GAN is trained on unpaired multi-contrast MR images, i.e., different contrasts are acquired from different subjects without requiring accurate registration. 
When training the network, two random training images $I_{{m}_i}, I_{m_j}$ in distinct contrasts ${m}_i , {m}_j  (i \neq j)$ are fed into Hyper-GAN as the source and target contrasts for updating network parameters.
For simplicity, we next present the training loss for the translation from  contrast ${m}_i$ to  ${m}_j$.

\noindent\textbf{Adversarial loss}~\cite{goodfellow2014}.
The hyper-encoder $E$ and hyper-decoder $G$ are required to generate a synthetic image $G( E( I_{m_i}, c_{i} ) , c_{j} )$ close to a real image, while the discriminator $D_{j}$ is to distinguish between this synthetic image $G( E( I_{m_i}, c_{i} ) , c_{j} )$ and a real image $I_{m_j}$.
The adversarial loss is defined as
\begin{equation}
	\mathcal{L}_{adv} (E, G, D_{j} ) = D_{j}  ( G( E( I_{m_i}, c_{i} ) , c_{j} )  )^2 +  ( 1- D_{j} ( I_{m_j} )  )^2 \ ,
\end{equation} 
where $E(\cdot, c_{i})$ and $G(\cdot, c_{j})$ respectively denote the outputs of hyper-encoder $E$ and hyper-decoder $G$ tuned by contrast codes $c_{i}$ and $c_{j}$. $G( E( I_{m_i}, c_{i} ),  c_{j} ) $ represents the synthetic $m_j$ contrast image translated from an $m_i$ contrast input.

\noindent\textbf{Cycle-consistency loss}~\cite{zhu2017}.
A cycle-consistency loss for $E$ and $G$ is to force the reconstructed image  $G ( E( G(  E( I_{{m}_i} , c_{i} ) , c_{j} ) , c_{j} ), c_{i} )$ (mapping from contrast  ${m}_i$ to ${m}_j$ and back) to be identical to the input $I_{{m}_i}$, which is written as
\begin{equation}
	\mathcal{L}_{cyc} (E, G) = \Vert G ( E( G(  E( I_{{m}_i} , c_{i} ) , c_{j} ) , c_{j} ), c_{i} )  - I_{{m}_i} \Vert _1 \ .
\end{equation}

\noindent\textbf{Common space loss.}
The hyper-encoder is adaptive to the contrast of input image, and we expect that  the extracted features by hyper-encoder should be in a common space shared by multi-contrast images for each subject, implicitly modeling the intrinsic tissue parameters of the subject. 
As paired training multi-contrast images are not given, we design  the following common space constraints. \\
(i) Identical loss.
As the images $I_{{m}_i}$ and $G(  E( I_{{m}_i} , c_{i} ) , c_{j} )$ are of the same subject but different contrasts, an identical loss is designed to enforce their extracted  features by hyper-encoder $E$ to be identical, which is written as
\begin{equation}
	\mathcal{L}_{id} (E,G) = \Vert E( G(  E( I_{{m}_i} , c_{i} ) , c_{j} ) , c_{j} )  - E(I_{{m}_i}, c_{i})  \Vert_1 \ .
\end{equation}
(ii) Reconstruction loss.
A reconstruction loss constrains that image $I_{{m}_i}$ could reconstruct itself by hyper-encoder and hyper-decoder, which is defined as
\begin{equation}
	\mathcal{L}_{rec} (E,G) = \Vert  G(  E( I_{{m}_i} , c_{i} ) , c_{i} )  -  I_{{m}_i} \Vert_1 \ .
\end{equation}
(iii) Contrast-classification loss.
Contrast-classification loss is to force classifier $C$ to predict the contrast of extracted  features by hyper-encoder $E$. We adversarially train the classifier to make deep features of multiple contrasts extracted by $E$ to be same distributed, i.e., within a common feature space, using a gradient reversal layer~\cite{ganin2015} in $C$, which flips gradient sign during backpropagation to force extracted deep features unable to be classified by $C$. 
This loss is defined as
\begin{equation}
\begin{aligned}
\mathcal{L}_{cla} (E,G,C) =
& L_{CE} ( C(E( G(  E( I_{{m}_i} , c_{i} ) , c_{j} ) , c_{j} )), c_{j} ) \\
& + L_{CE}( C(E( I_{{m}_i} , c_{i} )), c_{i} ) \ ,
\end{aligned}
\end{equation}
where $L_{CE}$ computes the cross entropy between estimated contrast probability by $C$ and real contrast code. Then the common space loss is defined as
\begin{equation}
\mathcal{L}_{com} (E,G,C) = \lambda_{id} \mathcal{L}_{id} + \lambda_{rec} \mathcal{L}_{rec} + \lambda_{cla} \mathcal{L}_{cla} \ .
\end{equation}
We heuristically set $\lambda_{id}$, $\lambda_{rec}$ and $\lambda_{cla}$ to $0.5$, $10$ and $0.2$  to make each term in a similar range of loss values as adversarial loss.

\noindent\textbf{Total loss.}
The total training loss of Hyper-GAN is summation of  above training losses over all training image pairs in distinct contrasts, which is defined as
\begin{equation}
	\mathcal{L}(E,G,C,D_{j}) = \mathcal{L}_{adv} + \lambda_{cyc} \mathcal{L}_{cyc} + \mathcal{L}_{com} \ ,
\end{equation}
where $\lambda_{cyc}$ is set to 10 as per~\cite{zhu2017}.
To optimize $\mathcal{L}$, the networks are divided into two groups, i.e., $\{ D_{j} \}_{j=1}^N$ and $\{ E,G,C \}$, which are alternately updated, and the networks are totally optimized in 100 epochs using an Adam optimizer with betas of $(0.5, 0.999)$. As in~\cite{zhu2017}, the learning rate is set to 0.0002 with a batch size of 1.  Our source code will be released on GitHub.

\section{Experiments}
\subsection{Data Sets}
\textbf{IXI dataset\footnote{https://brain-development.org/ixi-dataset/}.}
We utilize all 319 subjects from Guy's Hospital, and randomly split them into 150, 5 and 164 subjects for training, validation and testing.
Each subject contains three contrasts (T1w, T2w and PDw), and only one of three contrasts per subject is used for training  to generate unpaired  data. 
\begin{table}[!tp]
	\caption{Accuracies of different methods for arbitrary cross-contrast MR image translation on IXI dataset, which are averaged over test set and all 6 translation tasks. Each cell is formatted as ``mean (standard deviation)".}
	\label{tab:ixi_result}
	\centering
	\renewcommand\arraystretch{1.2}
	\begin{tabular}{l @{\hspace{0.45em}} l *{3}{@{\hspace{0.7em}} c}}
		\hline
		Method         & Ablated version & MAE             & PSNR         & SSIM          \\
		\hline
		CycleGAN       &                 & 0.0175 (0.0045) & 27.18 (2.17) & 0.857 (0.055)  \\
		StarGAN        &                 & 0.0174 (0.0052) & 27.77 (2.81) & 0.856 (0.068)  \\
		DGGAN          &                 & 0.0175 (0.0049) & 27.51 (2.59) & 0.858 (0.066) \\
		SUGAN          &                 & 0.0182 (0.0055) & 27.30 (2.88) & 0.845 (0.068)  \\
		ComboGAN       &                 & 0.0163 (0.0040) & 28.17 (2.34) & 0.876 (0.045)  \\
		\hdashline
		\multirow{6}{*}{Ours (CIN)}  
		& $\mathcal{L}_{adv \, + \, cyc}$                & 0.0145 (0.0039) & 28.89 (2.45)  & 0.897 (0.044)  \\
		& $\mathcal{L}_{adv \, + \, cyc+ \, id}$       & 0.0141 (0.0037) & 29.12 (2.46)  & 0.906 (0.040)  \\
		& $\mathcal{L}_{adv \, + \, cyc \, + \, rec}$ & 0.0144 (0.0043) & 29.09 (2.76)  & 0.900 (0.046)  \\
		& $\mathcal{L}_{adv \, + \, cyc \, + \, cla}$ &  0.0145 (0.0037) & 28.86 (2.32)  & 0.902 (0.038) \\
		& $\mathcal{L}_{adv \, + \, cyc \, + \, id \, + \, cla}$  & 0.0138 (0.0037) & 29.36 (2.54) & 0.909 (0.037)  \\
		& $\mathcal{L}_{adv \, + \, cyc \, + \, id \, + \, rec \, + \, cla}$  & 0.0138 (0.0043) & 29.45 (3.03) & \textbf{0.910} (\textbf{0.042}) \\
		\hdashline
		\multirow{6}{*}{Ours (FS)}  
		& $\mathcal{L}_{adv \, + \, cyc}$                 & 0.0154 (0.0032) & 28.28 (1.74) & 0.897 (0.030)  \\
		& $\mathcal{L}_{adv \, + \, cyc+ \, id}$        & 0.0141 (0.0040) & 29.24 (2.73) & 0.904 (0.041)  \\
		& $\mathcal{L}_{adv \, + \, cyc \, + \, rec}$  & 0.0140 (0.0047) & 29.39 (3.22) & 0.908 (0.045)  \\
		& $\mathcal{L}_{adv \, + \, cyc \, + \, cla}$  & 0.0141 (0.0036) & 29.04 (2.47) & 0.903 (0.041)  \\
		& $\mathcal{L}_{adv \, + \, cyc \, + \, id \, + \, cla}$  & 0.0135 (0.0033) & 29.49 (2.27) & \textbf{0.910} (\textbf{0.034})  \\
		& $\mathcal{L}_{adv \, + \, cyc \, + \, id \, + \, rec \, + \, cla}$  & \textbf{0.0133} (\textbf{0.0038}) & \textbf{29.64} (\textbf{2.70}) & \textbf{0.910} (\textbf{0.040})  \\
		\hline
	\end{tabular}
\end{table}

\noindent\textbf{BraTS 2019 dataset\footnote{https://www.med.upenn.edu/cbica/brats2019.html}.}
We use all 150 subjects from CBICA institution, and split them into 100, 5 and 45 subjects for  training, validation and testing.
Each subject contains four contrasts (T1w, T1Gd, T2w, FLAIR), and only one of four contrasts per subject is used for training. All volumes of both datasets are N4 corrected and peak normalized, and the intensities are linearly scaled to $[0,1]$.

\subsection{Experimental Results}
\label{sec:exp_result}
We compare our Hyper-GAN with the state-of-the-art unpaired multi-contrast image translation methods, including  StarGAN~\cite{choi2018}, DGGAN~\cite{tang2018dual}, SUGAN~\cite{sohail2019unpaired} and ComboGAN~\cite{anoosheh2018}, for arbitrary cross-contrast MR image translation.
Our Hyper-GAN includes two versions, i.e., the contrast modulator using filter scaling (FS) or conditional instance normalization (CIN).
All  experiments are performed on 2D saggital slices.
Quantitatively, we compute  mean absolute error (MAE), peak signal-to-noise ratio (PSNR), and structural similarity (SSIM) between  3D volumes of ground truth and image translation results.
The final accuracy is averaged over  test set and all translation tasks between all possible source/target contrast pairs, i.e., 6 tasks for IXI  and 12 tasks for BraTS 2019.
A paired two-sided Wilcoxon signed-rank test is conducted to compare the performance.
\begin{figure}[!tp]
	\includegraphics[width=\textwidth]{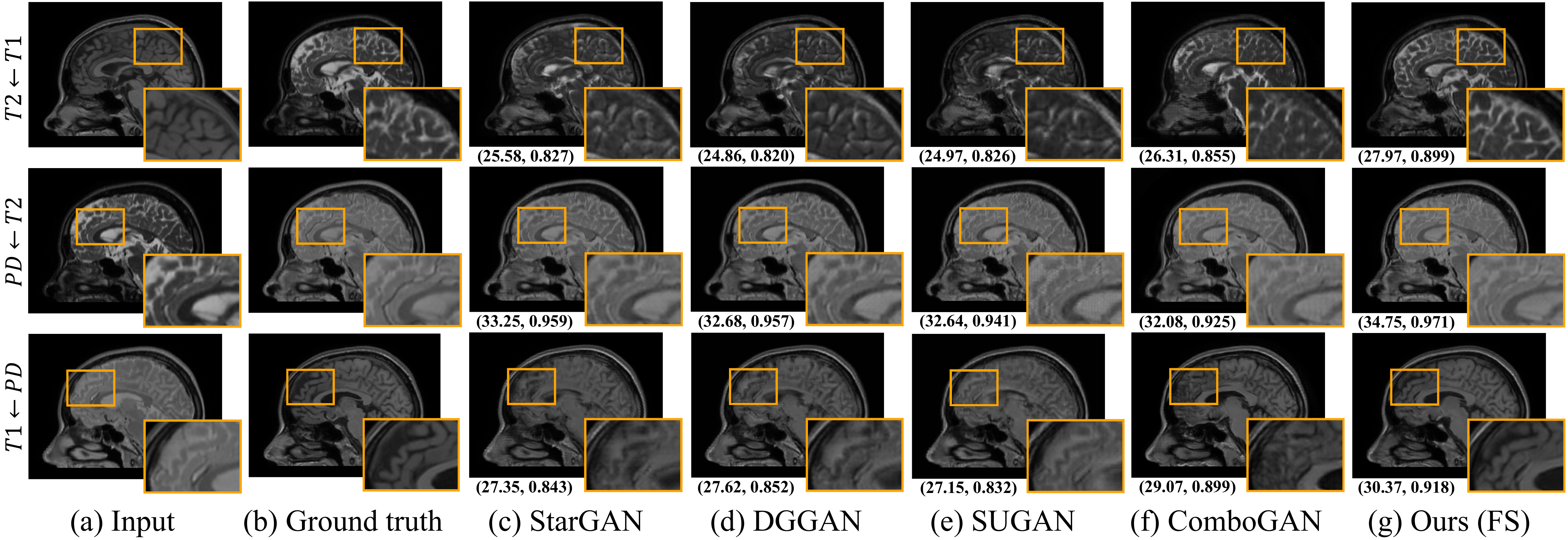}
	\centering
	\caption{Visual comparison of cross-contrast MR image translation results on IXI dataset. Top to bottom: T1w-to-T2w, T2w-to-PDw and PDw-to-T1w translation results on a test subject. The values under each sub-image are PSNR and SSIM scores.}
	\label{Fig:ixi-visual}
\end{figure}

\noindent\textbf{Results on IXI dataset.}
Table~\ref{tab:ixi_result} reports the results on IXI dataset of three contrasts.
It shows that, compared with StarGAN, DGGAN, SUGAN and ComboGAN, our Hyper-GAN (i.e., $\mathcal{L}_{adv \, + \, cyc \, + \, id \, + \, rec \, + \, cla}$) achieves significantly better performance in all metrics ($p < .001$) and produces 29.45/29.64 (using CIN/FS respectively) in PSNR, comparing favorably with 27.77 of StarGAN, 27.51 of DGGAN, 27.30 of SUGAN and 28.17 of ComboGAN. 
Specifically, our method achieves the highest accuracies for all 6 translation tasks in all metrics.

\noindent\textbf{Effectiveness of training losses.} 
In Table~\ref{tab:ixi_result}, we also compare Hyper-GAN trained with different losses defined in Sect.~\ref{sec:train_loss}.
Interestingly, even our baseline (i.e., $\mathcal{L}_{adv \, + \, cyc}$) performs better than compared methods ($p < .005$) and obtains 28.89/28.28 in PSNR, justifying effectiveness of our network design with hyper-encoder/hyper-decoder using contrast modulators. 
Starting from this baseline, each extra loss enforcing  common feature space constraint improves results in all metrics except that contrast-classification loss in CIN version produces higher SSIM but comparable MAE and PSNR scores.
Specifically, the identical and reconstruction losses respectively improve  PSNR from 28.89/28.28 of baseline to 29.12/29.24 and 29.09/29.39.
In addition, the PSNR scores are further improved to 29.36/29.49 when both identical and contrast-classification losses are utilized and our whole model performs best. This indicates the effectiveness of our defined common  space loss implicitly modeling  the shared underlying anatomical structures of  each subject.
Both FS and CIN versions achieve good performance while FS works generally better.
Figure~\ref{Fig:ixi-visual} shows visual comparison results.

\begin{figure}[!tp]
	\includegraphics[width=\textwidth]{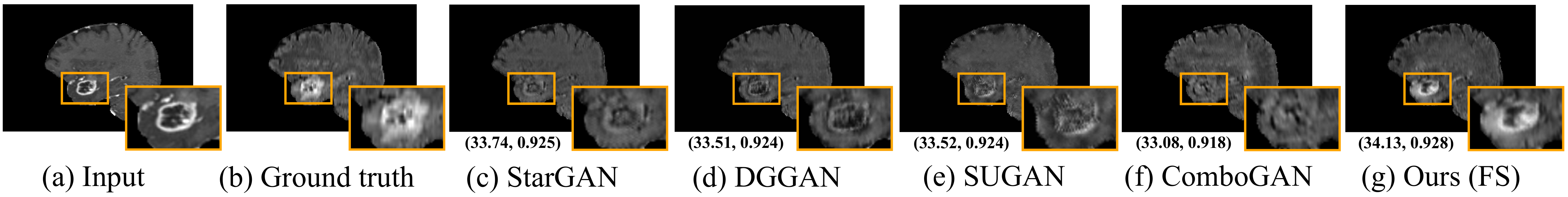}
	\caption{Visual comparison of   T1Gd-to-FLAIR image translation results on BraTS 2019 dataset. The values under each sub-image are PSNR and SSIM scores.}
	\label{Fig:brats-visual}
\end{figure}

\noindent\textbf{Results on BraTS 2019 dataset.}
Table~\ref{tab:brats_result} reports the results on BraTS 2019 dataset of four contrasts. Our Hyper-GAN obtains significantly higher accuracies than compared methods in all metrics ($p < .001$) and achieves 31.66/31.90 (using CIN/FS respectively) in PSNR, comparing favorably with 30.75 of StarGAN, 30.81 of DGGAN, 30.37 of SUGAN and 30.55 of ComboGAN.
Specifically, our method works better than all compared methods for all 12 tasks in all metrics, except MAE score of StarGAN for FLAIR-to-T2 translation (0.0103 of StarGAN vs. 0.0105 of ours).
Figure~\ref{Fig:brats-visual} shows the T1Gd-to-FLAIR image translation results, and our method performs best even with lesions.

\begin{table}[!tp]
	\caption{Accuracies of different methods for arbitrary cross-contrast MR image translation on BraTS 2019 dataset, which are averaged over test set and all 12 translation tasks. Each cell is formatted as ``mean (standard deviation)".}
	\label{tab:brats_result}
	\centering
	\renewcommand\arraystretch{1.2}
	\begin{tabular}{l*{4}{@{\hspace{1.2em}} c}}
		\hline
		Method      &  MAE             &  PSNR            &  SSIM              &  \# of parameters  \\
		\hline
		StarGAN     & 0.0083 (0.0029) &  30.75 (2.64)  &  0.905 (0.028)  & 50.72M \\
		DGGAN       & 0.0081 (0.0026) &  30.81 (2.46)  &  0.908 (0.023)  & 58.74M \\
		SUGAN       & 0.0087 (0.0021) &  30.37 (1.97)  &  0.909 (0.013)  & 82.63M \\
		ComboGAN    & 0.0081 (0.0025) &  30.55 (2.17)  &  0.913 (0.015)  & 71.02M \\
		\hdashline
		Ours (CIN)  & 0.0072 (0.0022) &  31.66 (2.46)  &  0.922 (0.018)  & 23.71M \\
		Ours (FS)   & \textbf{0.0070} (\textbf{0.0026})   & \textbf{31.90} (\textbf{2.74})  & \textbf{0.930} (\textbf{0.018})  & 22.57M \\
		\hline
	\end{tabular}
\end{table}

\noindent\textbf{Comparison with CycleGAN.} 
We also compare our Hyper-GAN with CycleGAN~\cite{zhu2017} in Table~\ref{tab:ixi_result}.
Note that concatenation of backbones of hyper-encoder and hyper-decoder is identical to generator in CycleGAN as discussed in Sect.~\ref{sec:net_arch}. 
As CycleGAN is for one-to-one contrast translation, we train $\frac{N\times(N-1)}{2}$ CycleGAN models for a N-contrast translation task, each of which is for a specific pair of contrasts. We average the results over different contrast pairs as final accuracy of CycleGAN.
The results show that our Hyper-GAN significantly outperforms CycleGAN ($p < .005$), demonstrating the effectiveness of our network design based on contrast modulator and common feature space constraint.

\noindent\textbf{Computational Efficiency.} 
As shown in Table~\ref{tab:brats_result}, our parameter size is  2 and 3 times smaller than StarGAN/DGGAN and SUGAN/ComboGAN respectively. 
Also for BraTS 2019 dataset, CycleGAN requires 169.56M parameters that is 7 times larger than ours.
This is because Hyper-GAN utilizes only a single pair of hyper-encoder and hyper-decoder, while the number of encoders/decoders in ComboGAN or CycleGAN scales linearly or quadratically with contrast number.

\section{Conclusion}
%
%

We have proposed a unified GAN for unpaired multi-contrast MR image translation. 
It can flexibly translate between different contrast pairs using a unified network consisting of hyper-encoder/hyper-decoder and common feature space constraint. 
This  design enables the network to fully investigate the common anatomical structures by common feature space modeling, and contrast-specific imaging by  hyper-network design of hyper-encoder and hyper-decoder. 
It achieves state-of-the-art performance, and outperforms  previous GAN-based image translation models that depend on multiple encoders/decoders for different contrast pairs.
In the future, we are interested in the extension to multi-contrast, multi-institute setting~\cite{dewey2020disentangled}, as well as the combination with segmentation task~\cite{jiang2020unified,yuan2019,zhou2020brain}.

\subsubsection*{Acknowledgments.}
This work is supported by the NSFC (12026603, 12026605, 12090021, 61721002, 11690011, U1811461, U20B2075, 11971373).

\bibliographystyle{splncs04}
\bibliography{paper335}

\begin{thebibliography}{10}
\providecommand{\url}[1]{\texttt{#1}}
\providecommand{\urlprefix}{URL }
\providecommand{\doi}[1]{https://doi.org/#1}

\bibitem{alharbi2019}
Alharbi, Y., Smith, N., Wonka, P.: Latent filter scaling for multimodal
  unsupervised image-to-image translation. In: IEEE Conference on Computer
  Vision and Pattern Recognition. pp. 1458--1466 (2019)

\bibitem{anoosheh2018}
Anoosheh, A., Agustsson, E., Timofte, R., Van~Gool, L.: Combogan: Unrestrained
  scalability for image domain translation. In: IEEE Conference on Computer
  Vision and Pattern Recognition Workshops. pp. 783--790 (2018)

\bibitem{bernstein2004}
Bernstein, M.A., King, K.F., Zhou, X.J.: Handbook of {MRI} pulse sequences.
  Elsevier (2004)

\bibitem{bui2020flow}
Bui, T.D., Nguyen, M., Le, N., Luu, K.: Flow-based deformation guidance for
  unpaired multi-contrast {MRI} image-to-image translation. In: Martel, A.L.,
  Abolmaesumi, P., Stoyanov, D., et~al. (eds.) MICCAI 2020. pp. 728--737.
  Springer, Cham (2020). \doi{10.1007/978-3-030-59713-9\_70}

\bibitem{choi2018}
Choi, Y., Choi, M., Kim, M., Ha, J.W., Kim, S., Choo, J.: Stargan: Unified
  generative adversarial networks for multi-domain image-to-image translation.
  In: IEEE Conference on Computer Vision and Pattern Recognition. pp.
  8789--8797 (2018)

\bibitem{dar2019}
Dar, S.U., Yurt, M., Karacan, L., Erdem, A., Erdem, E., {\c{C}}ukur, T.: Image
  synthesis in multi-contrast {MRI} with conditional generative adversarial
  networks. IEEE Transactions on Medical Imaging  \textbf{38}(10),  2375--2388
  (2019)

\bibitem{dewey2020disentangled}
Dewey, B.E., Zuo, L., Carass, A., He, Y., Liu, Y., Mowry, E.M., et~al.: A
  disentangled latent space for cross-site {MRI} harmonization. In: Martel,
  A.L., Abolmaesumi, P., Stoyanov, D., et~al. (eds.) MICCAI 2020. pp. 720--729.
  Springer, Cham (2020). \doi{10.1007/978-3-030-59728-3\_70}

\bibitem{dumoulin2016}
Dumoulin, V., Shlens, J., Kudlur, M.: A learned representation for artistic
  style. In: International Conference on Learning Representations (2017)

\bibitem{ganin2015}
Ganin, Y., Lempitsky, V.: Unsupervised domain adaptation by backpropagation.
  In: International Conference on Machine Learning. pp. 1180--1189 (2015)

\bibitem{goodfellow2014}
Goodfellow, I., Pouget-Abadie, J., Mirza, M., Xu, B., Warde-Farley, D., Ozair,
  S., Courville, A., Bengio, Y.: Generative adversarial nets. In: Advances in
  Neural Information Processing Systems. pp. 2672--2680 (2014)

\bibitem{ha2017}
Ha, D., Dai, A.M., Le, Q.V.: Hypernetworks. In: International Conference on
  Learning Representations (2017)

\bibitem{huang2017}
Huang, Y., Shao, L., Frangi, A.F.: {DOTE}: {D}ual c{O}nvolutional fil{T}er
  l{E}arning for super-resolution and cross-modality synthesis in {MRI}. In:
  Descoteaux, M., Maier-Hein, L., Franz, A., et~al. (eds.) MICCAI 2017. pp.
  89--98. Springer, Cham (2017). \doi{10.1007/978-3-319-66179-7\_11}

\bibitem{huang2020mcmt}
Huang, Y., Zheng, F., Cong, R., Huang, W., Scott, M.R., Shao, L.: {MCMT-GAN}:
  Multi-task coherent modality transferable {GAN} for 3{D} brain image
  synthesis. IEEE Transactions on Image Processing  \textbf{29},  8187--8198
  (2020)

\bibitem{hui2018}
Hui, L., Li, X., Chen, J., He, H., Yang, J.: Unsupervised multi-domain image
  translation with domain-specific encoders/decoders. In: International
  Conference on Pattern Recognition. pp. 2044--2049 (2018)

\bibitem{isola2017}
Isola, P., Zhu, J.Y., Zhou, T., Efros, A.A.: Image-to-image translation with
  conditional adversarial networks. In: IEEE Conference on Computer Vision and
  Pattern Recognition. pp. 1125--1134 (2017)

\bibitem{jiang2020unified}
Jiang, J., Veeraraghavan, H.: Unified cross-modality feature disentangler for
  unsupervised multi-domain {MRI} abdomen organs segmentation. In: Martel,
  A.L., Abolmaesumi, P., Stoyanov, D., et~al. (eds.) MICCAI 2020. pp. 347--358.
  Springer, Cham (2020). \doi{10.1007/978-3-030-59713-9\_34}

\bibitem{jog2017}
Jog, A., Carass, A., Roy, S., Pham, D.L., Prince, J.L.: Random forest
  regression for magnetic resonance image synthesis. Medical Image Analysis
  \textbf{35},  475--488 (2017)

\bibitem{liu2021dual}
Liu, X., Xing, F., Prince, J.L., Carass, A., Stone, M., Fakhri, G.E., Woo, J.:
  Dual-cycle constrained bijective {VAE-GAN} for tagged-to-cine magnetic
  resonance image synthesis. In: International Symposium on Biomedical Imaging
  (2021)

\bibitem{long2015}
Long, J., Shelhamer, E., Darrell, T.: Fully convolutional networks for semantic
  segmentation. In: IEEE Conference on Computer Vision and Pattern Recognition.
  pp. 3431--3440 (2015)

\bibitem{roy2016}
Roy, S., Chou, Y.Y., Jog, A., Butman, J.A., Pham, D.L.: Patch based synthesis
  of whole head {MR} images: Application to {EPI} distortion correction. In:
  International Workshop on Simulation and Synthesis in Medical Imaging. pp.
  146--156 (2016)

\bibitem{sohail2019unpaired}
Sohail, M., Riaz, M.N., Wu, J., Long, C., Li, S.: Unpaired multi-contrast {MR}
  image synthesis using generative adversarial networks. In: International
  Workshop on Simulation and Synthesis in Medical Imaging. pp. 22--31 (2019)

\bibitem{tang2018dual}
Tang, H., Xu, D., Wang, W., Yan, Y., Sebe, N.: Dual generator generative
  adversarial networks for multi-domain image-to-image translation. In: Asian
  Conference on Computer Vision. pp. 3--21 (2018)

\bibitem{vranic2019}
Vranic, J., Cross, N., Wang, Y., Hippe, D., de~Weerdt, E., Mossa-Basha, M.:
  Compressed sensing--sensitivity encoding ({CS-SENSE}) accelerated brain
  imaging: reduced scan time without reduced image quality. American Journal of
  Neuroradiology  \textbf{40}(1),  92--98 (2019)

\bibitem{yu2019}
Yu, B., Zhou, L., Wang, L., Shi, Y., Fripp, J., Bourgeat, P.: {Ea-GANs}:
  edge-aware generative adversarial networks for cross-modality {MR} image
  synthesis. IEEE Transactions on Medical Imaging  \textbf{38}(7),  1750--1762
  (2019)

\bibitem{yu2020sample}
Yu, B., Zhou, L., Wang, L., Shi, Y., Fripp, J., Bourgeat, P.: Sample-adaptive
  {GANs}: Linking global and local mappings for cross-modality {MR} image
  synthesis. IEEE Transactions on Medical Imaging  \textbf{39}(7),  2339--2350
  (2020)

\bibitem{yuan2019}
Yuan, W., Wei, J., Wang, J., Ma, Q., Tasdizen, T.: Unified attentional
  generative adversarial network for brain tumor segmentation from multimodal
  unpaired images. In: Shen, D., Liu, T., Peters, T.M., et~al. (eds.) MICCAI
  2019. pp. 229--237. Springer, Cham (2019).
  \doi{10.1007/978-3-030-32248-9\_26}

\bibitem{zaitsev2015}
Zaitsev, M., Maclaren, J., Herbst, M.: Motion artifacts in {MRI}: A complex
  problem with many partial solutions. Journal of Magnetic Resonance Imaging
  \textbf{42}(4),  887--901 (2015)

\bibitem{zhou2020brain}
Zhou, T., Canu, S., Vera, P., Ruan, S.: Brain tumor segmentation with missing
  modalities via latent multi-source correlation representation. In: Martel,
  A.L., Abolmaesumi, P., Stoyanov, D., et~al. (eds.) MICCAI 2020. pp. 533--541.
  Springer, Cham (2020). \doi{10.1007/978-3-030-59719-1\_52}

\bibitem{zhu2017}
Zhu, J.Y., Park, T., Isola, P., Efros, A.A.: Unpaired image-to-image
  translation using cycle-consistent adversarial networks. In: IEEE
  International Conference on Computer Vision. pp. 2223--2232 (2017)

\end{thebibliography}

\end{document}